\documentclass[journal]{IEEEtran}
\usepackage{bbding}
\usepackage{multirow}
\usepackage{xcolor}
\usepackage{hyperref}

\ifCLASSINFOpdf
   \usepackage[pdftex]{graphicx}
\else
\fi
\begin{document}
%
\title{Raw Image Deblurring}
%
%
%


\author{Chih-Hung Liang,
        Yu-An Chen,
        Yueh-Cheng Liu,
        and~Winston H. Hsu,~\IEEEmembership{Senior Member,~IEEE}%
        \thanks{All authors are affiliated with National Taiwan University}%
        \thanks{Email: \{r06922057, r07922076, liu115\}@cmlab.csie.ntu.edu.tw; whsu@ntu.edu.tw}%
        }

%
%

\markboth{IEEE TRANSACTIONS ON MULTIMEDIA
}%
{Liang \MakeLowercase{\textit{et al.}}: IEEE TRANSACTIONS ON MULTIMEDIA}
%



\maketitle

\begin{abstract}
Deep learning-based blind image deblurring plays an essential role in solving image blur since all existing kernels are limited in modeling the real world blur. Thus far, researchers focus on powerful models to handle the deblurring problem and achieve decent results. For this work, in a new aspect, we discover the great opportunity for image enhancement (e.g.,  deblurring) directly from RAW images and investigate novel neural network structures benefiting RAW-based learning. However, to the best of our knowledge, there is no available RAW image deblurring dataset. Therefore, we built a new dataset containing both RAW images and processed sRGB images and design a new model to utilize the unique characteristics of RAW images. The proposed deblurring model,  trained solely from RAW images, achieves the state-of-art performance and outweighs those trained on processed sRGB images. Furthermore, with fine-tuning, the proposed model, trained on our new dataset, can generalize to other sensors. Additionally, by a series of experiments, we demonstrate that existing deblurring models can also be improved by training on the RAW images in our new dataset. Ultimately, we show a new venue for further opportunities based on the devised novel raw-based deblurring method and the brand-new Deblur-RAW dataset.
\end{abstract}


%
\IEEEpeerreviewmaketitle

%
%
%
%

 




\section{Introduction}
\IEEEPARstart{B}{lur} is mainly caused by accumulating optical signals captured by the sensor during the exposure time. It usually occurs when the camera is shaking and/or objects in captured scenes are moving. Deblurring task aiming to restore sharp images has attracted researchers for attending the needs of growing hand-held camera users and supporting various computer vision tasks such as object detection and image segmentation.

\begin{figure}[t]
\centering
\includegraphics[width=1\linewidth]{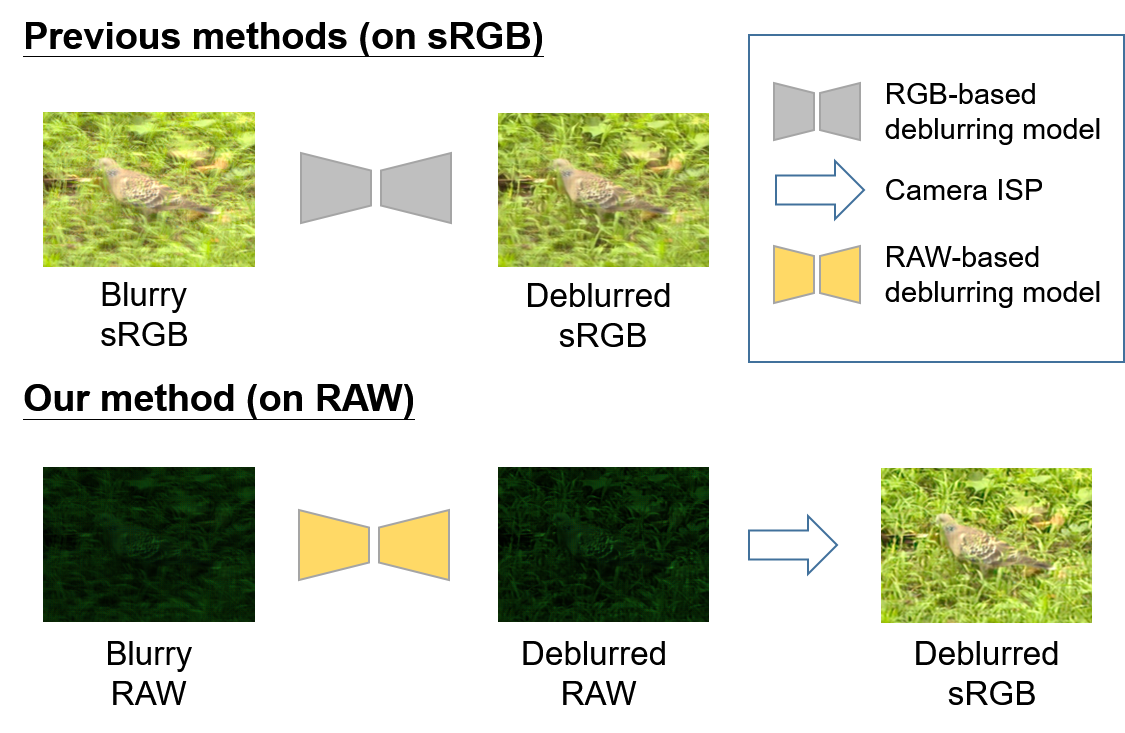}
\caption{To address the image deblurring problem, previous methods mainly focus on models trained on processed sRGB datasets while we promote a novel view of solving the deblurring problem by training on RAW images. Furthermore, a novel network structure is proposed in this work to meet the particular properties of RAW images. The complementary effect of the RAW images and the proposed network structure leads to better performance on the deblurring task. [Best viewed in color.]}
\label{fig:fig1}
\end{figure}

Image deblurring is still a highly ill-posed problem. Conventional methods usually make some assumptions or constraints to model the complicated blur kernel \cite{sun2013edge,michaeli2014blind,hyun2013dynamic,hyun2014segmentation}. However, these methods cannot directly apply to real-world cases because the approximations are still inaccurate. In recent years, \cite{nah2017deep,su2017deep} try to generate pairs of blur and sharp images in a blind way. They record high frame rate videos and then synthesize blurry images by averaging successive frames. Their methods relieve us of sophisticated blur kernels designing and make related researches flourish. However, the averaged RGB values still cannot simulate the real-world blur well since they have been processed by camera image processing pipeline (IPP). The non-linear steps in IPP, such as demosaicing, white balance, color correction, and gamma compression may affect the generated sRGB images and make them non-linear to real sensor data. It also makes the prior dataset \cite{nah2017deep} require post-processing to reduce the domain gap. The issues in the RGB-based datasets highlight the necessity and importance of the RAW-based dataset for image deblurring.


In this work, we create a new dataset Deblur-RAW to make deblurring on raw sensor data possible. We capture RAW videos in various scenes with a fixed camera setting and then split them into successive RAW frames. Inspired by \cite{nah2017deep,su2017deep}, we generate blurred RAW images blindly by averaging successive RAW frames (3-5 frames) and use the center one as the sharp ground truths. As directly operating on raw sensor data, we can generate more realistic blurred images without any post-processing such as gamma correction \cite{nah2017deep}.

In addition to collecting the dataset, we also propose a novel network architecture that aims to properly utilize special properties in RAW images. Unlike sRGB images, RAW images are stored with color filters, such as Bayer pattern, and X-trans. Each position in images stores not only spatial information but also specific color sensor value. Prior works \cite{chen2018learning,xu2019towards,zhang2019zoom} try to pack the same color into one channel and treat RAW images as four channels (RGBG) images. While the packing strategy is intuitive, it may make images lose spatial order which is important for deblurring. Therefore, we propose a novel two-branch network architecture, one branch focuses on spatial structure and the other one focuses on information of the same color sensors. Furthermore, we introduce bidirectional cross-modal attention (BCA) between the two branches to make them jointly enhance each other.

In the experiment section, we demonstrate that directly deblurring on RAW images is beneficial. Because of more information kept in RAW images, we can restore sharper details and structure. Additionally, after applying our proposed novel network architecture, we get more performance gains. It implies that the designed network is more suitable for RAW images. We outperform the state-of-the-art RGB-based image-deblurring methods, both quantitatively and qualitatively. In brief, our contributions can be summarized as follows:

\begin{itemize}
\item We introduce Deblur-RAW, the first RAW image deblurring dataset. We generate RAW blurred images by averaging successive RAW frames and use center one as the sharp ground truths.
\item We demonstrate that directly deblurring on high-bit RAW images helps the restoration of image details and structure. It makes us outperform the RGB-based state-of-the-art methods.
\item We propose a novel network architecture to utilize special properties of RAW images. We design a two-branch network to handle spatial and colored sensor information respectively. Furthermore, we introduce BCA to make two branches be jointly enhanced by each other.
\end{itemize}

\section{Related Work}

Most of the deblurring approaches are based on the following blur model formulation. \cite{gong2017motion,sun2015learning}
\begin{equation}
{I_{B} = K(M) * I_{S} + N}
\end{equation}
where $I_{B}$ is the blurry image, $K(M)$ are the blur kernels depended on motion field $M$, $I_{S}$ is the latent sharp image and $N$ is the noise. The deblurring problems can be defined as a non-blind or blind way depending on whether information about blur or blur kernel ($K(M)$) is available or not. 
Most conventional works focus on non-blind deblurring. They try to model the blind blur kernels with simple assumptions and priors. Kim et al. \cite{hyun2013dynamic} proposed a segmentation-based method that segments a blurry image and estimates the non-uniform blur kernel in each segment. Kim and Lee \cite{hyun2014segmentation} approximated the pixel-wise blur kernel as locally linear. Michaeli et al. \cite{michaeli2014blind} restored sharp images by patch-based priors in down-scaled images. Yu et al. \cite{yu2014efficient} proposed a patch-wise non-uniform deblurring algorithm to estimate each kernel locally and employed the total variation regularization to recover a latent image. However, in practice, blur kernels are usually unknown.

In recent years, learning-based methods have demonstrated promising results in various computer vision tasks such as super-resolution \cite{ledig2017photo,lai2017deep,lim2017enhanced,tai2017image,caballero2017real, yang2018drfn}, denoising \cite{zhang2017beyond,liao2017multi}, object removal \cite{cai2018semantic,chen2018robust}, style transfer \cite{luan2017deep,isola2017image,zhu2017unpaired}, and image deblurring \cite{kupyn2018deblurgan,nah2017deep,zhang2018dynamic}. Some image deblurring methods try to estimate the blur kernels by deep learning methods, and then restore latent sharp images \cite{xu2014deep,sun2015learning,gong2017motion}. Xu et al. \cite{xu2014deep} proposed a deblurring method based on the blur kernels that can be decomposed into a small number of filters. Sun et al. \cite{sun2015learning} tried to estimate the probabilities of the predefined motion kernels for each image patch by CNN and then restored the latent image by optimization method. Gong et al. \cite{gong2017motion} estimated the motion flow by the fully convolutional network, and then recovered the sharp image via non-blind deconvolution. These methods utilize deep learning to get more accurate blur kernel estimation. However, they usually fail in real-world cases as simple kernel assumptions cannot properly model the non-uniform and complicated blur kernels.

As blur kernels in real-world images are usually complicated and unknown, some works tried to restore sharp images directly in a blind manner \cite{nah2017deep,su2017deep,kupyn2018deblurgan,zhang2018dynamic,tao2018scale,zhang2019deep,lu2019unsupervised}. Nah et al. \cite{nah2017deep} adopted kernel-free methods in both dataset generation and latent image estimation. They proposed a state-of-the-art method by using the multi-scale network and released an image deblurring dataset which is widely used in later works. Su et al. \cite{su2017deep} were inspired by Nah et al \cite{nah2017deep} and also released a video deblurring dataset which consists of 71 video sequences. They proposed an intuitive but effective video deblurring method which restored frames by stacking successive blurry frames. Kupyn et al. \cite{kupyn2018deblurgan} restored sharp images by using the GAN-based model. Tao et al. \cite{tao2018scale} enhanced the multi-scales method with the recurrent structure which enables the network to share weights across scales. Rather than deblurring on different scales, Zhang et al. \cite{zhang2019deep} choose to deblur on small patches cropped from whole images and achieve the state-of-the-art performance on GoPro dataset. Lu et al. \cite{lu2019unsupervised} try to deblur in an unsupervised manner. Inspired by CycleGAN \cite{zhu2017unpaired}, they use the two-branch model to learn blur and deblur from unpaired data. These methods mainly focus on deblurring on the sRGB domain as there are only datasets with sRGB images. They lose rich and valuable information which can be gained from raw sensor data.

Some prior works tried to address the blurry artifact by the information from raw sensor data \cite{zhen2013enhanced, trimeche2005multichannel}. Zhen et al. \cite{zhen2013enhanced} utilized the information from inertial sensors to restore the RAW images. They built a digital image system to acquire RAW images in conjunction with 3-axis acceleration data. With the acceleration data, the camera motion and the blur kernel can be estimated. Then the blur kernel is applied in a MAP estimation to restore the RAW image. Trimeche et al. \cite{trimeche2005multichannel} proposed a novel multi-channel image restoration algorithm to reduce the optical blur caused by the camera optical system. They applied the modified iterative Landweber algorithm combined with the adaptive denoising technique to each color channel separately on the RAW images. To enhance the robustness of the iterative process, they used the adaptive filter based on the local polynomial approximation of neighboring pixels from dynamically selected windows. Besides, to avoid false coloring due to independent channel filtering in RGB space, they also propose a novel saturation control mechanism to attenuate the iterative restoration in near-saturated regions. While these works benefited from the RAW data, they are still non-blind methods and based on assumptions. Furthermore, they mainly focus on the blur caused by the camera and still struggle for the cases caused by object motion.



There are prior works that have shown that raw sensor data can enhance image processing tasks \cite{plotz2017benchmarking,chen2018learning,schwartz2018deepisp,xu2019towards,zhang2019zoom}. 
Pl\"{o}tz et al. \cite{plotz2017benchmarking} collected Darmstadt Noise Dataset, a new benchmark dataset for RAW image denoising, which greatly inspired related researches.
Schwartz et al. \cite{schwartz2018deepisp} presented DeepISP, a full end-to-end deep neural model of the camera images signal processing pipeline.
Chen et al. \cite{chen2018learning} addressed the extremely low-light image enhancement problem by learning from raw sensor data.
Xu et al. \cite{xu2019towards} used RAW data to help details and structure restoration for super-resolution and achieve better performance in real scenarios.
Zhang et al. \cite{zhang2019zoom} introduced SR-RAW, a new RAW image dataset for super-resolution. They collected the dataset via camera optical zoom. They also demonstrated that directly operating on raw sensor data is indeed beneficial.
All these works showed that high-bit raw sensor data is beneficial to various image processing tasks. However, to the best of our knowledge, there is no available RAW image dataset for image deblurring, which restricts the prior methods to the sRGB domain.

As a result, we create a new RAW image dataset, Deblur-RAW, for image deblurring. Besides, we find that packing strategy used by prior works may break spatial order and is not suitable for image deblurring. To address the issue, we propose a novel network architecture which can jointly consider both spatial structure and color sensor information. By directly operating on raw sensor data with the proposed network, our method can restore more details and achieve better performance than the prior state-of-the-art RGB-based methods.

\section{Deblur-RAW Dataset}
\label{section:dataset}

\begin{figure}[t]
\centering
\includegraphics[width=1\linewidth]{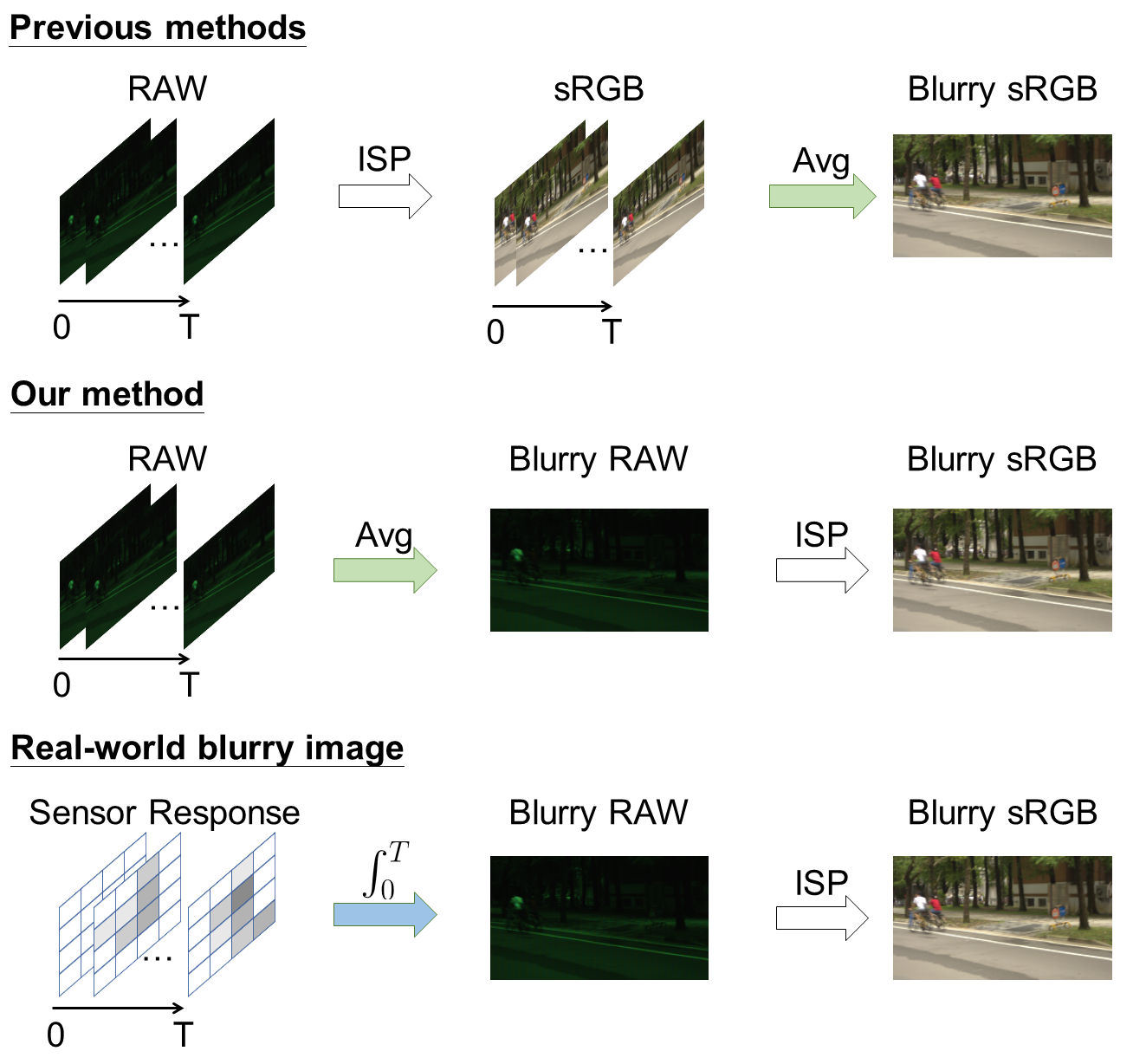}
\caption{Real-world blur is caused by accumulating signals captured by camera during the exposure time. To simulate the real-world blur, previous methods take an average of processed sRGB images which are non-linear to original sensor data. In contrast, we directly operate on RAW images, which enables us to generate more realistic blurred images without post-processing. [Best viewed in color.]
}
\label{fig:dataset}
\end{figure}

To perform end-to-end learning for RAW image deblurring, we collect a new data, Deblur-RAW, containing pairs of blur and sharp RAW images and their processed sRGB images.
Blurry images are mainly caused by accumulating signals captured by camera sensors during the exposure time. Blur accumulation process can be formulated as follow \cite{nah2017deep}:
\begin{equation}
B_{RAW} = \frac{1}{T}\int_{t=0}^{T}S(t)dt \ \simeq \  \frac{1}{M}\sum_{i=0}^{M}S[i] 
\end{equation}
where T and S(t) are exposure time and signals captured by camera sensor at time t respectively. Likewise, M, S[i] are the number of recorded frames and the i-th sharp RAW frame in the recorded video. Inspired by Nah et al. \cite{nah2017deep} and Su et al. \cite{su2017deep}, we generate blurred RAW images blindly by averaging successive sharp RAW frames recorded by the camera. As directly operating on RAW images, we can generate realistic blurred images without the influences of image processing pipeline and any post-processing such as gamma correction \cite{nah2017deep}. The comparison of the dataset generation pipeline is shown in Figure \ref{fig:dataset}.


We collect our dataset by Canon EOS 6D, EF 17-40mm f/4L USM. With the help of Magic Lantern, an open-source enhancement for canon cameras, we are able to record RAW videos that contain successive RAW frames. We set the camera shutter speed at 1/250 to 1/400 second depending on the lightness to make sure each recorded frame is sharp. Moreover, to make all the frames contain enough lightness, we fix the camera aperture at the largest value f/4.0. As the limitation of camera write-out time, we can only record RAW videos with 30 fps. Then we split the RAW videos into RAW frames and average varying number (3-5) of successive frames to generate blurred RAW images. And we take the center frame of the averaged ones as the sharp ground truth. Because all the frames are taken from the same RAW videos, they share almost the same metadata except the name and time. As a result, we use the metadata of the sharp ground truths as the metadata of the generated blurred RAW images. Besides, we also provide sRGB images of the generated RAW pairs by LibRaw, an open-source library, with the default setting.
There are a few main steps in the image processing pipeline of LibRaw. First, RAW images are scaled by the daylight white balance. Then, Adaptive Homogeneity-Directed (AHD) is used to perform demosaicing. The AHD selects the direction of interpolation to maximize a homogeneity metric, thus typically minimizing color artifacts. After demosaicing, the color space conversion is performed to transform the RGB value in the image into a standardized device-independent color space, sRGB. Finally, the gamma correction with 2.222 power and 4.5 slope is applied to generate the final results.

\begin{figure}[t]
\begin{center}
\includegraphics[width=1\linewidth]{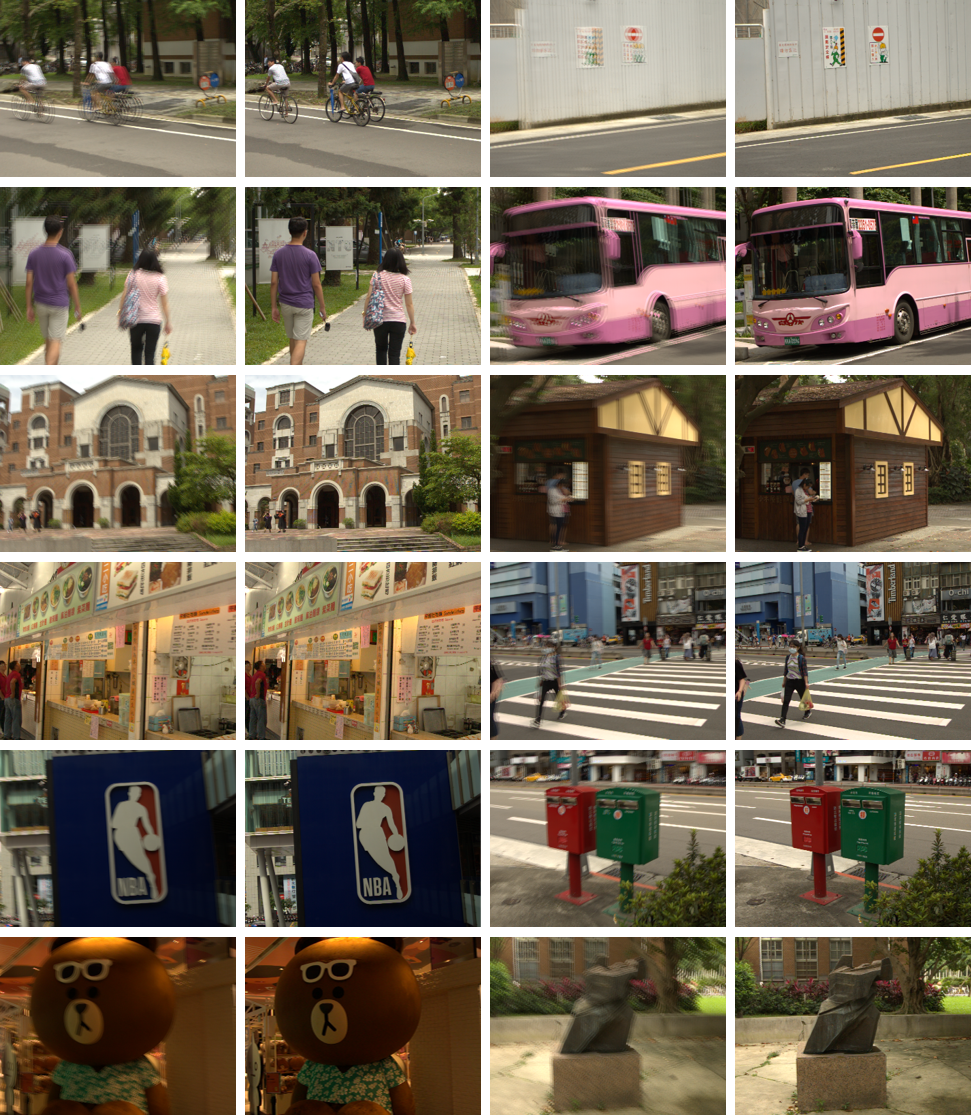}
\end{center}
\caption{Some collected cases in Deblur-RAW dataset. As RAW images are hard to be observed, only the generated sRGB images are demonstrated. [Best viewed in color.]
}
\label{fig:dataset_samples}
\end{figure}

\begin{table}[t]
\centering
\caption{Image deblurring dataset comparison.
}
\begin{tabular}{lcccc}
\hline
Dataset & \multicolumn{1}{c}{sRGB} & \multicolumn{1}{c}{Blindly} & \multicolumn{1}{c}{RAW} & pairs \\ \hline
DeblurGAN \cite{kupyn2018deblurgan} & \CheckmarkBold & -              & -              & 1151 \\
GoPro    \cite{nah2017deep}         & \CheckmarkBold & \CheckmarkBold & -              & 3214 \\
Su et al. \cite{su2017deep}         & \CheckmarkBold & \CheckmarkBold & -              & 6708 \\
Deblur-RAW (ours)                   & \CheckmarkBold & \CheckmarkBold & \CheckmarkBold & \textbf{10252} \\
\hline
\end{tabular}
\label{table:dataset_comparison}
\end{table}

We record RAW videos at various scenes, such as garden, school, shopping mall, intersection, sports field, playground, and so on. Some cases are shown in Figure \ref{fig:dataset_samples}. There are 103 RAW videos, 10252 generated RAW images pairs, and processed sRGB images in the Deblur-RAW dataset, which is much more than the previous image deblurring datasets proposed by \cite{nah2017deep} and \cite{su2017deep}. The comparison with the widely used datasets is shown in Table \ref{table:dataset_comparison}. The proposed Deblur-RAW dataset enables us to learn an end-to-end deblurring model on RAW images and enhances prior RGB-based methods. We will also release our dataset and look forward to inspiring further researches. The dataset is available in the following link: \textcolor{magenta}{\url{https://github.com/bob831009/raw_image_deblurring}}.

\section{Proposed Method}
\label{section:proposed_method}
In this section, we provide every detail of our method especially the intimate relationship between the proposed method and the training data--RAW images. RAW images store sensor values filtered by a color filter array (CFA), such as Bayer filter or X-trans filter, where each pixel contains both spatial structure and color sensor information. The previous training strategy is to separate RAW images into four different channels where pixels with the same color are packed into a channel. However, this packing strategy breaks the spatial order of RAW images and leads to information loss during the deblurring process. To address the issue, novel network architecture with the ability to utilize both spatial order and color sensor information is proposed in this work. As a result, the performance of our proposed model, which trained on RAW images, is superior to models that trained on processed sRGB images.

\begin{figure*}[t]
\centering
\includegraphics[width=0.8\linewidth]{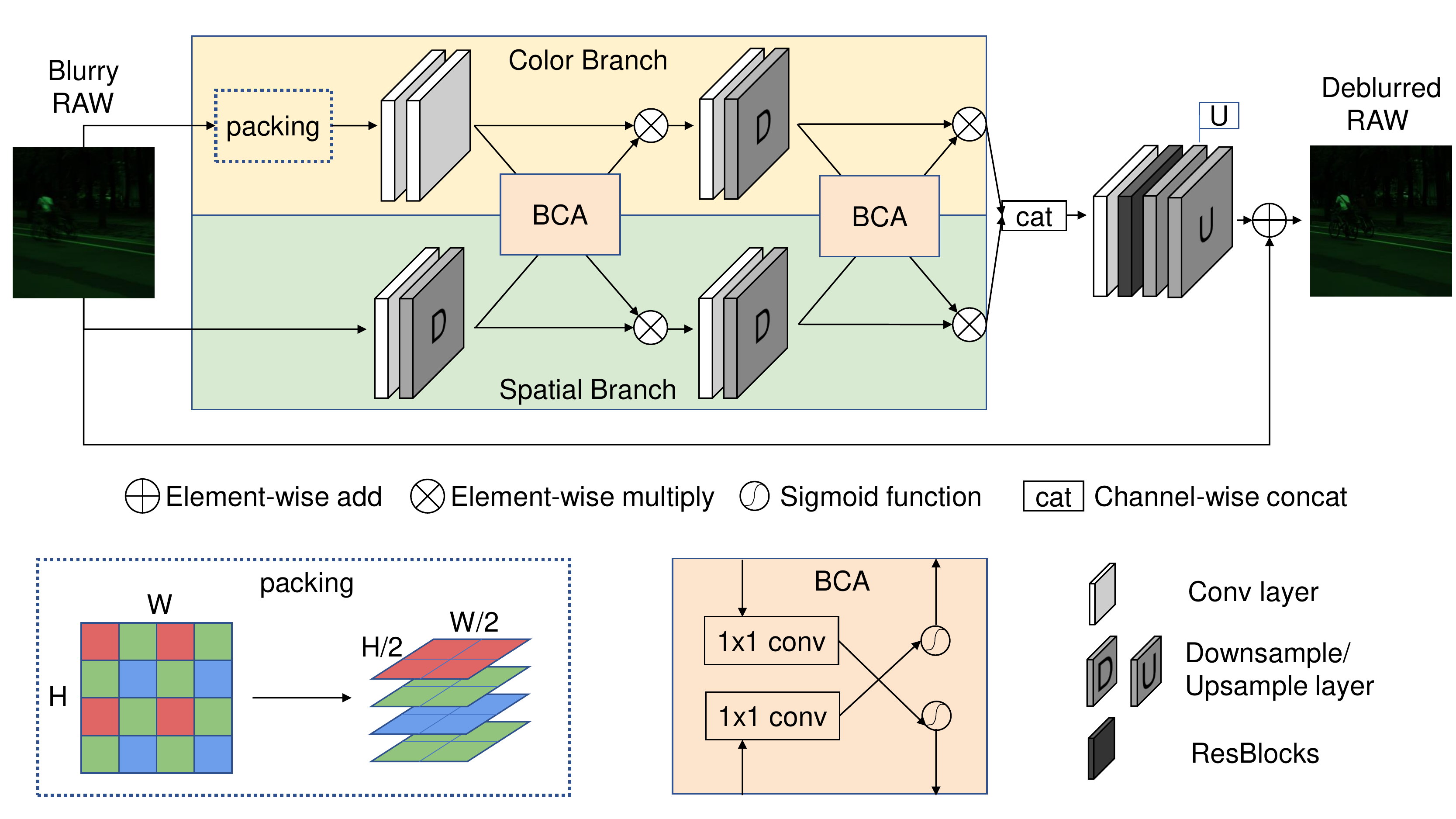}
\caption{\textbf{Overall network architecture:} Our network is organized by three key components (1) The spatial branch encoder is in charge of the spatial structure of input RAW images. (2) The color branch encoder is the place where packed RGBG images are processed. Follow by the packing strategy, pixels with the same color filter are put into the same channel so that the color branch is able to handle information from different color filters separately. (3) The bidirectional cross-modal attention (BCA) enables the two branches mentioned above to enhance each other. [Best viewed in color.]}
\label{fig:network_structure}
\end{figure*}

\subsection{Network Architecture}


Our overall network architecture is shown in Figure \ref{fig:network_structure} and Table \ref{table:network_architecture}. The encoder-decoder network is adopted as the main backbone and nine ResBlocks are placed between the encoder and decoder network to help the learning process. A skip connection is added between input and model output which enable our model to focus on leaning differences between blurry and sharp RAW images. The two-branch encoder is employed to utilize raw sensor information without losing spatial structure. It’s notable that, one of the two branches is in charge of spatial structure of raw images while another one focuses on color sensor information. Moreover, motivated by Yang et al. \cite{Yang_2019_ICCV}, bidirectional cross-modal attention (BCA) is introduced and set between two branches which enables them to be enhanced by each other.

\subsubsection{Spatial and Color Encoder}


To learn from RAW sensor data, most of the existing works treat RAW images as four-channel (RGBG) images where pixels with the same color are packed into the same channel. However, RAW images store sensor values that contain both spatial structure and color sensor information. This packing strategy may downgrade the image resolution and breaks the spatial order of RAW images. Therefore, to make use of two different properties of RAW images mentioned above, the two-branch encoder is adopted. The branch that in charge of the spatial structure has whole RAW images as the input while the one focuses on color sensor information has packed RAW images as the input. In other words, the input of the spatial encoder preserves the original spatial information of the RAW image, and the input of the color encoder is aligned by color. In the experiment section, we demonstrate that the designed architecture indeed helps us gain more meaningful information from RAW images for deblurring.

\subsubsection{Bidirectional Cross-modal Attention}
\begin{figure*}[t]
\centering
\includegraphics[width=0.78\linewidth]{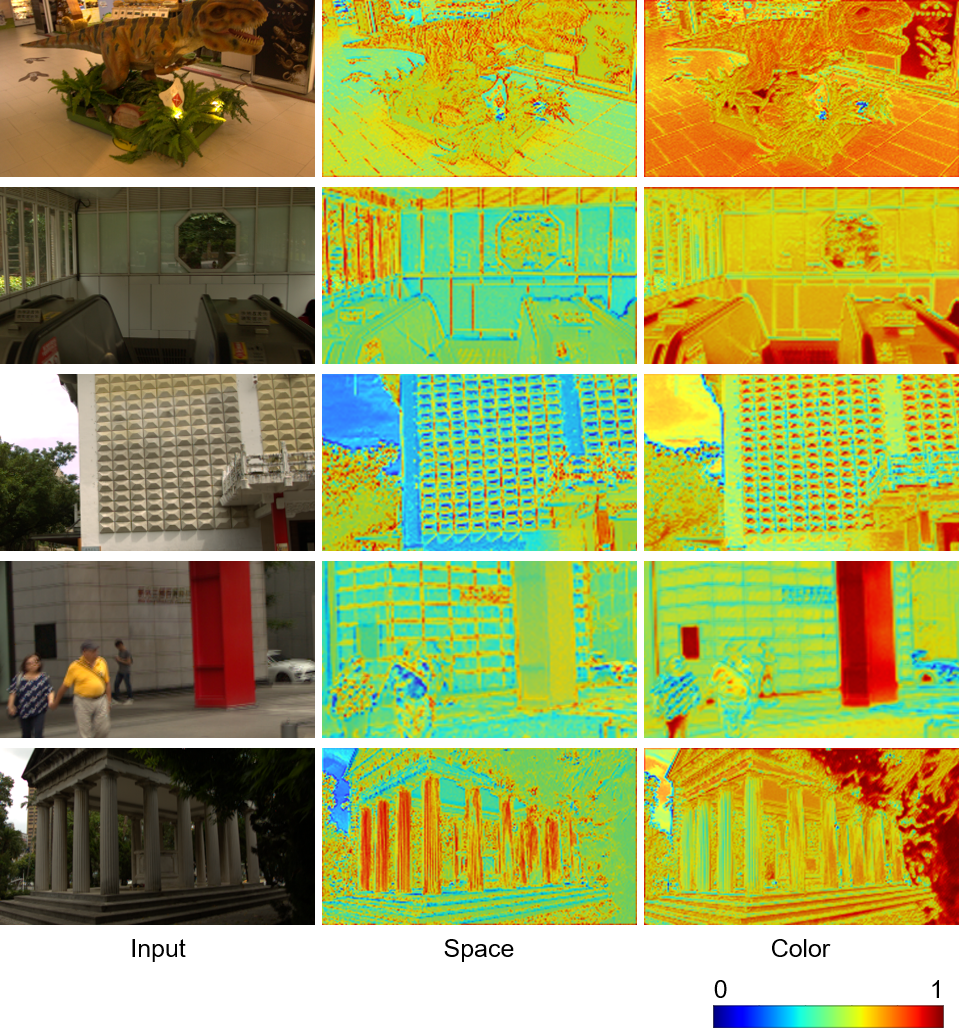}
\caption{\textbf{Visualization of attentive feature maps from BCA}. We can observe that the attention maps of the spatial branch mainly focus on silhouettes and edges, and those of the color branch focus on interior regions. Two branches are helpful and complementary to each other, which enables us to properly utilize RAW images for deblurring. [Best viewed in color.]
}
\label{fig:bca_attention}
\end{figure*}

Inspired by Yang et al. Bidirectional Cross-modal Attention (BCA) is adopted between the spatial and color branches so that two branches are able to enhance each other.
The BCA can be formulated as following:
\begin{equation}
\hat{M}^{space} = M^{space} \otimes Sigmoid(Conv_{1*1}(M^{color}))
\label{eq:color2space}
\end{equation}
\begin{equation}
\hat{M}^{color} = M^{color} \otimes Sigmoid(Conv_{1*1}(M^{space}))
\label{eq:space2color}
\end{equation}
$M^{space}$ and $M^{color}$ are the extracted feature map of spatial and color branch respectively. $\otimes$ denotes element-wise multiplication. There are two attention directions, one is space to color, and another is color to space. The attention weights are generated by the feature maps of the other branch. With the attention weights, the feature maps can be further enhanced by the information from the other branch. In the direction from color to space, as formulated in equation \ref{eq:color2space}, the spatial branch can also consider the information from the same color filter. In the direction from space to color, as formulated in equation \ref{eq:space2color}, the color branch can also keep the spatial structure of original images. We add BCA at each downsampling layer in our encoder network, which makes two branches be gradually enhanced by each other from low-level to high-level features. By the visualization of the attention weights of two branches in Figure \ref{fig:bca_attention}, we can observe that the spatial branch mainly focuses on silhouettes and edges while the color branch focuses on interior regions. Two branches are complementary to each other. In the experiment section, we will show that the BCA indeed benefits the RAW image deblurring process and performance is improved significantly as well.



\begin{table*}[t]
\centering
\caption{The detail architecture of the proposed network.
}
\begin{tabular}{|c|c|c|c|c|c|c|c|}
\hline
\multicolumn{3}{|c|}{\textbf{Layer}} & \textbf{Channel} & \textbf{Kernal size} & \textbf{Stride} & \textbf{Padding} & \textbf{Size of Output} \\ \hline
\multirow{4}{*}{Spatial Branch} & Input Image &  & - & - & - & - & 128 * 128 * 1 \\ \cline{2-8} 
 & Input Conv & Conv2d, BatchNorm, ReLU & 64 & 7 & 1 & 3 & 128 * 128 * 64 \\ \cline{2-8} 
 & DownSampling 1 & Conv2d, BatchNorm, ReLU & 128 & 3 & 2 & 1 & 64 * 64 * 128 \\ \cline{2-8} 
 & DownSampling 2 & Conv2d, BatchNorm, ReLU & 256 & 3 & 2 & 1 & 32 * 32 * 256 \\ \hline
\multirow{4}{*}{Color Branch} & Input Stack Image &  & - & - & - & - & 64 * 64 * 4 \\ \cline{2-8} 
 & Input Conv & Conv2d, BatchNorm, ReLU & 64 & 3 & 1 & 1 & 64 * 64 * 64 \\ \cline{2-8} 
 & DownSampling 1 & Conv2d, BatchNorm, ReLU & 128 & 3 & 1 & 1 & 64 * 64 * 128 \\ \cline{2-8} 
 & DownSampling 2 & Conv2d, BatchNorm, ReLU & 256 & 3 & 2 & 1 & 32 * 32 * 256 \\ \hline
\multirow{4}{*}{BCA} & \multirow{2}{*}{BCA1} & Conv2d, Sigmoid & 128 & 1 & 1 & 0 & 64 * 64 * 128 \\ \cline{3-8} 
 &  & Conv2d, Sigmoid & 128 & 1 & 1 & 0 & 64 * 64 * 128 \\ \cline{2-8} 
 & \multirow{2}{*}{BCA2} & Conv2d, Sigmoid & 256 & 1 & 1 & 0 & 32 * 32 * 256 \\ \cline{3-8} 
 &  & Conv2d, Sigmoid & 256 & 1 & 1 & 0 & 32 * 32 * 256 \\ \hline
Concat Layer & Concat Layer & Conv2d, BatchNorm, ReLU & 256 & 3 & 1 & 1 & 32 * 32 *256 \\ \hline
ResBlocks & ResBlock * 9 & Conv2d, BatchNorm & 256 & 3 & 1 & 1 & 32 * 32 *256 \\ \hline
\multirow{2}{*}{Decoder} & UpSampling 2 & ConvTranspose2d, BatchNorm, ReLU & 128 & 3 & 2 & 1 & 64 * 64 * 128 \\ \cline{2-8} 
 & UpSampling 1 & ConvTranspose2d, BatchNorm, ReLU & 64 & 3 & 2 & 1 & 128 * 128 * 64 \\ \hline
Output Layer & Output Layer & Conv2d, Tanh & 1 & 7 & 1 & 3 & 128 * 128 * 1 \\ \hline
\end{tabular}
\label{table:network_architecture}
\end{table*}

\subsection{Loss Functions}
There are two loss functions which are used in our models, $L2$ loss and SSIM loss. $L2$ loss, shown in equation \ref{eq:l2_loss}, enables the model to focus on the pixel-wise difference.
\begin{equation}
L_{mse} = \frac{1}{N} \sum_{i=1}^{N} \left \| I_{pred} - I_{gt} \right \|_{2}^{2}
\label{eq:l2_loss}
\end{equation}
Different from $L2$ loss which only focuses on pixel-wise similarity, SSIM loss encourages model to consider the structure similarity of a group of pixels and generate visually pleasing images. SSIM for pixel $p$ is defined as:
\begin{equation}
SSIM(p) = \frac{(2\mu _{x}\mu _{y} + c_{1})(2\sigma _{x,y} + c_{2}) }{(\mu _{x}^2 + \mu_{y}^2 + c_{1})(\sigma_{x}^2 + \sigma_{y}^2 + c_{2})}
\label{eq:ssim}
\end{equation}
where $\mu_{x}$ and $\mu_{y}$ are mean and $\sigma_{x}$ and $\sigma_{y}$ are standard deviation of the predicted and ground truth RAW image. $\sigma_{x,y}$ is the covariance. By averaging the SSIM of each pixel, SSIM loss is denoted as:
\begin{equation}
L_{ssim} = \sum_{p \in P}^{ }1 - SSIM(p)
\label{eq:ssim_loss}
\end{equation}
$L2$ loss is stable in early training stage, but it makes the model be prone to produce blurry results. In contrast, SSIM loss is instable in early training stage, but it makes the model generate visually pleasing images. As a result, we combine both by a constant $\lambda$ as our final loss function. Our final loss function is defined as:
\begin{equation}
L = L_{mse} + \lambda L_{ssim}
\label{eq:total_loss}
\end{equation}

\subsection{Implementation Details}
We provide the details of our model hyperparameters and training proceduce in this section. First, we subtract the black level value from RAW images and divide them by pixel maximum value to normalize into [0, 1]. As we directly operating on RAW images, we need to perform augmentation without breaking the color filter pattern. Therefore, there is few augmentation we can do. We augment the input RAW images by randomly cropping into 256*256 patches. We use Adam as our optimization function. We fix the learning rate at 0.0001 at first 500 epochs and linearly decay to 0 in the following 500 epochs. All the results reported in this paper are trained with 1000 epochs, which takes about one days on single NVIDIA TESLA V100 GPU. As our model is fully convolutional, the resolution of the input images is only restricted by GPU memory resource.

\section{Experimental Results}
\label{section:experiments}
In this section, we show the superiority of the proposed dataset and method by a series of experiments. For objective evaluations, all the methods are tested on the same hardware environment. We fine-tune the prior RGB-based methods with the processed sRGB images in Deblur-RAW. Since our method directly deblurs on RAW images, we also process the restored RAW images to the sRGB domain by the same image processing pipeline, LibRaw, for a fair comparison. We use peak signal-to-noise ratio (PSNR) and structure similarity index (SSIM) as the evaluation metrics. All the experiments are shown in the following sections.

\subsection{Quantitative Results}
We compare our method with some representative image deblurring methods: Nah et al. \cite{nah2017deep}, which releases GoPro dataset and deblurs with the multi-scale network; Kupyn et al. \cite{kupyn2018deblurgan}, a GAN-based deblurring method; Tao et al. \cite{tao2018scale}, which enhances the multi-scale network by recurrent structure; Lu et al. \cite{lu2019unsupervised}, which learns the deblurring network in an unsupervised manner; and Zhang et al. \cite{zhang2019deep}, which deblurs on several cropped patches and achieves outstanding performance in GoPro dataset without high computational cost. For all the methods, we use the public released weights and fine-tune on the sRGB images in Deblur-RAW following their setup. The RAW images restored by our method are also processed to the sRGB domain by the same image processing pipeline mentioned in Section \ref{section:dataset} for an objective evaluation.

\begin{table}[t]
\centering
\caption{Quantitative comparison with the state-of-the-art image deblurring methods.
}
\begin{tabular}{|c|c|c|c|}
\hline
 & PSNR $\uparrow$ & SSIM $\uparrow$ & Runtime (s) \\ \hline
Nah et al. \cite{nah2017deep}            & 27.85            & 0.8803            & 1.769 \\ \hline
DeblurGAN \cite{kupyn2018deblurgan}      & 26.58            & 0.8519            & 0.007 \\ \hline
Tao et al. \cite{tao2018scale}           & 28.69            & 0.9246            & 0.335 \\ \hline
Lu et al.  \cite{lu2019unsupervised}     & 24.60            & 0.8110            & 0.029 \\ \hline
DMPHN\_1\_2\_4\_8 \cite{zhang2019deep}   & 28.73            & 0.9071            & 0.079 \\ \hline
SDNet4 \cite{zhang2019deep}              & 29.24            & 0.9195            & 0.169 \\ \hline
our method                               & \textbf{29.80}   & \textbf{0.9285}   & 0.014 \\ \hline
\end{tabular}
\label{table:quantitative}
\end{table}

The evaluation results are shown in Table \ref{table:quantitative}. It is noted that our method outperforms the state-of-the-art image deblurring methods without high computational cost. Directly deblurring on raw sensor data differentiates our method from the state-of-the-arts. As the low-bit sRGB images are processed by the image processing pipeline, they lose the meaningful information which can be gained from high-bit raw sensor data. By our collected Deblur-RAW dataset, we are able to learn and deblur on RAW images directly, which enables our method to achieve better performance than the prior RGB-based state-of-the-arts.

\subsection{Qualitative results}
\begin{figure*}[t]
\centering
\includegraphics[width=0.9\linewidth]{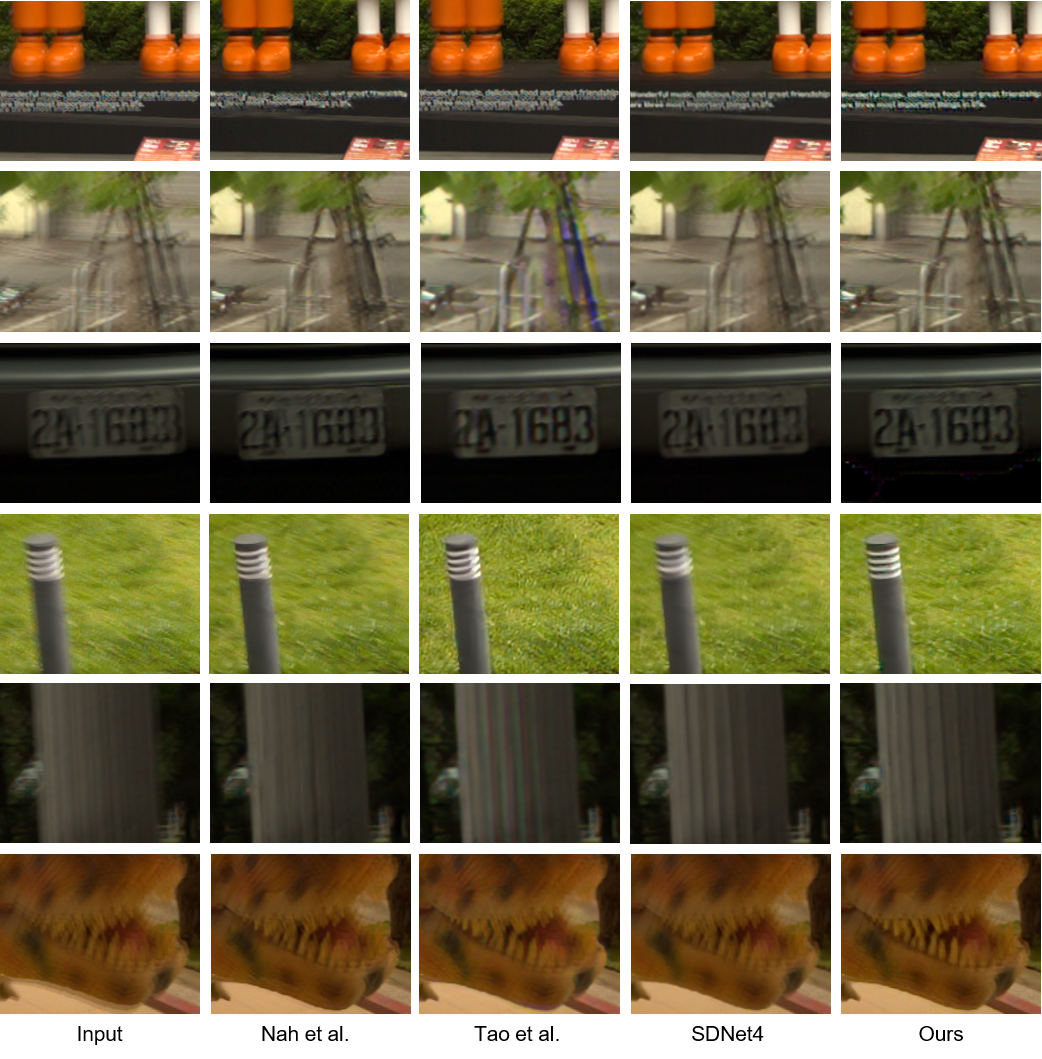}
\caption{Qualitative evaluation with the RGB-based state-of-the-art methods. By leveraging rich and valuable information in RAW images, our method is able to restore sharper details and structure. [Best viewed in color.]}
\label{fig:qualitative_evaluation}
\end{figure*}

\begin{figure*}[t]
\centering
\includegraphics[width=0.9\linewidth]{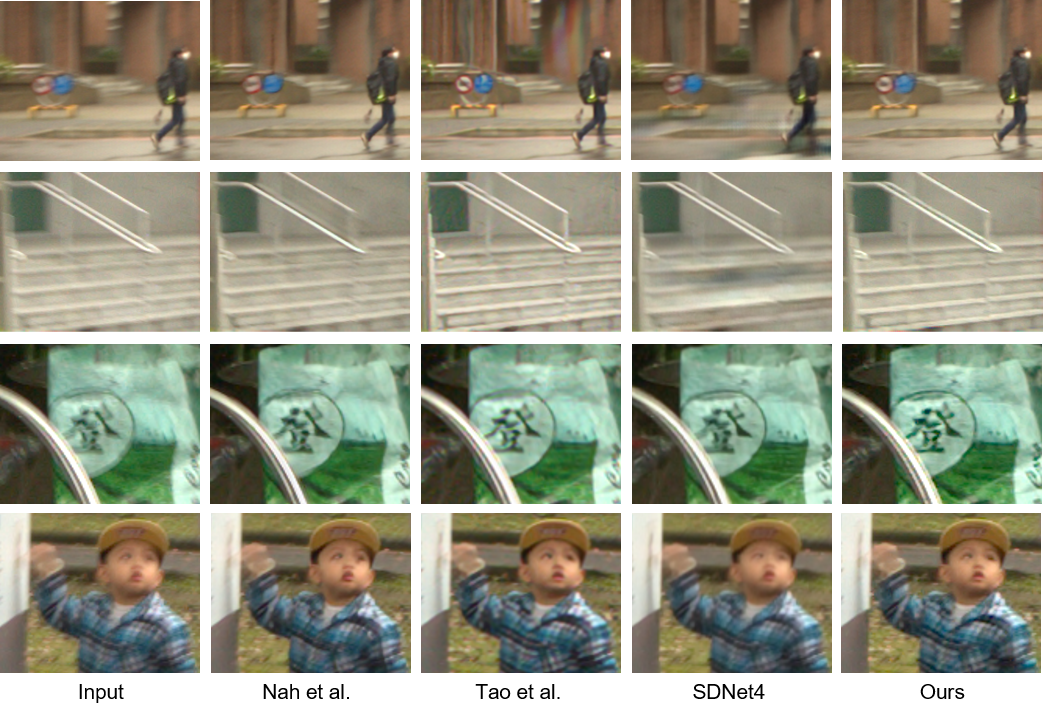}
\caption{Qualitative evaluation with the RGB-based state-of-the-art methods on real-world images. We can observe that our method performs better not only on the synthetic data but also on the real-world images. [Best viewed in color.]}
\label{fig:qualitative_evaluation_real_world}
\end{figure*}

Besides the numerical analysis, we also compare the quality of the images restored by the state-of-the-art methods and ours, shown in Figure \ref{fig:qualitative_evaluation}. In addition to the synthetic cases, the results on real blurry images captured by a long exposure (1/10 sec) camera are also provided in Figure \ref{fig:qualitative_evaluation_real_world}. We show the comparisons with \cite{nah2017deep,tao2018scale,zhang2019deep}, which achieve better performance in Table \ref{table:quantitative}. Because of the reduced information in processed sRGB images, limited details that the state-of-the-art methods can restore. In contrast, ours can restore images with clean structure and fine details. It indicates that rich and valuable information kept in RAW images is indeed beneficial to the image deblurring task.


\subsection{Ablation Study}
We conduct intensive ablation studies to verify that each designed component and loss function discussed in Section \ref{section:proposed_method} are helpful. For a fair comparison, all the models are trained with the same setting and configuration. Besides evaluating final processed sRGB images, we also show the evaluation of RAW images. The results are shown in Table \ref{table:ablation_study}.
We can observe that the model trained with L2 and SSIM loss performs better than those without combination.
Besides, it is noted that the model with only the spatial encoder performs much better than the one with only the color encoder.
It indicates that the packing strategy which breaks the spatial order and downgrades the resolution of original images is unfavorable to image deblurring task. The model only trained with packed RAW images may not perform as well as the one trained with whole RAW images.

Furthermore, we can also observe that the design of the two-branch encoder makes us get more performance gain. It indicates that the information from the same color filter is beneficial to the restoration of RAW images. Moreover, after adding Bidirectional Cross-modal Attention (BCA), we obtain the best performance. It means that the interactively learning of spatial and color branches is truly helpful. By progressively fusing the feature maps extracted from two branches, both of them can be enhanced by each other.

Besides, to verify the performance gain comes from the designed two-branch architecture rather than introducing more parameters, two experiments about the Color and Spatial encoder variant are also conducted. We double the channel of each layer in the single-branch models. It is noted that the single-branch model with more parameters still cannot outperform our two-branch version. It demonstrates that the proposed two-branch architecture extracts more meaningful information from RAW images and is beneficial to image deblurring.

The results of our ablation study demonstrate that all designed components are important and influential. Based on our ablation study, we choose the model with all the designed components as our final version which enables us to effectively restore sharp images from raw sensor data.

\begin{table*}[t]
\centering
\caption{Ablation Study of the proposed method.
}
\begin{tabular}{l|c|c|c|c|c}
\hline
                                       & \multicolumn{2}{c|}{RAW}          & \multicolumn{2}{c|}{sRGB}          & Runtime \\
Methods                                & PSNR $\uparrow$ & SSIM $\uparrow$ & PSNR $\uparrow$ & SSIM $\uparrow$ & (s)     \\ \hline
ours - Color encoder, L2 loss           & 40.81           & 0.9834          & 28.13           & 0.8962          & 0.007   \\
ours - Color encoder, SSIM loss         & 41.19           & 0.9854          & 28.51           & 0.9061          & 0.007   \\
ours - Color encoder, L2 and SSIM loss  & 41.50           & 0.9860          & 28.79           & 0.9109          & 0.007   \\
ours - Color encoder w/o black level     & 41.31           & 0.9856          & 28.75           & 0.9112          & 0.013   \\
ours - Spatial encoder w/o black level     & 41.89           & 0.9869          & 29.30           & 0.9180          & 0.013   \\
ours - Color encoder w/o black level 2x channel     & 41.50           & 0.9860          & 28.93           & 0.9141          & 0.016   \\
ours - Spatial encoder w/o black level 2x channel    & 42.11           & 0.9872          & 29.41           & 0.9213          & 0.016   \\
ours - Color and Spatial encoders        & 42.34           & 0.9880          & 29.57           & 0.9241          & 0.014   \\
ours - Color and Spatial encoders w/ BCA & \textbf{42.71}  & \textbf{0.9888} & \textbf{29.80}  & \textbf{0.9285} & 0.014   \\ \hline
\end{tabular}
\label{table:ablation_study}
\end{table*}

\subsection{Generalization to other methods}
To demonstrate that raw sensor data is beneficial to the deblurring task, we also train DMPHN\_1\_2\_4\_8 \cite{zhang2019deep}, the state-of-the-art image deblurring model, with our collected RAW images. We choose DMPHN\_1\_2\_4\_8 because RAW images cannot perform scaling directly and it deblurs on several cropped patches rather than on multi-scale images. Furthermore, it achieves better performance on GoPro dataset against prior methods without high computational cost. As a result, we choose DMPHN\_1\_2\_4\_8 as our target experimental method. We follow the same setting in \cite{zhang2019deep}, except the channel number of the input and output layer. We also follow the experimental setting described in Section \ref{section:experiments}, processing the restored RAW images to the sRGB domain for a fair comparison. Furthermore, we add our designed key components into the network to verify that all of them are beneficial for RAW image deblurring. All the models deblur on cropped patches with a maximum proper size for network prediction and then tile the patches into the final images. The evaluation results are shown in Table \ref{table:generalize_other_method}.

It is noted that the model trained with RAW images achieves better performance without other modifications. In indicates that valuable information kept in high-bit RAW images benefits the image deblurring process. Additionally, we can also observe that the model gradually performs better after adding the designed key components. The increasingly improving performances indicate that the designed components are helpful to RAW image deblurring and are complementary to other methods.

\begin{table*}[t]
\centering
\caption{Generalization to the other method.
}
\begin{tabular}{l|c|c|c|c|c}
\hline
                                            & \multicolumn{2}{c|}{RAW}          & \multicolumn{2}{c|}{sRGB}          & Runtime \\
Methods                                     & PSNR $\uparrow$ & SSIM $\uparrow$ & PSNR $\uparrow$ & SSIM $\uparrow$ & (s)     \\ \hline
DMPHN - sRGB                    & -               & -               & 28.73           & 0.9071          & 0.079    \\
DMPHN - RAW                     & 41.68           & 0.9856          & 28.98           & 0.9055          & 0.051    \\
DMPHN - RAW two branches         & 42.00           & 0.9871          & 29.07           & 0.9147          & 0.088    \\
DMPHN - RAW two branches w/ BCA  & \textbf{42.71}  & \textbf{0.9885} & \textbf{29.40}  & \textbf{0.9214} & 0.099    \\ \hline
\end{tabular}
\label{table:generalize_other_method}
\end{table*}

\begin{figure*}[ht]
\centering
\includegraphics[width=0.85\linewidth]{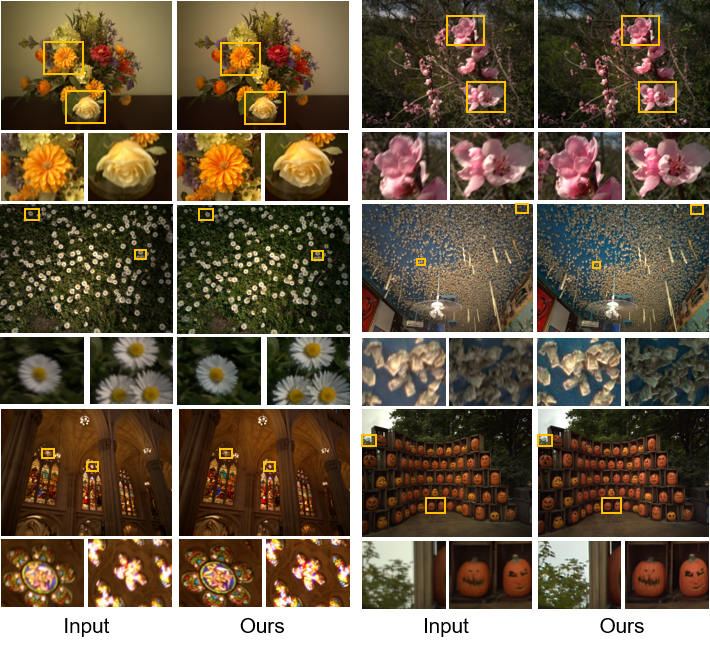}
\caption{\textbf{Qualitative results on HDR+ dataset.} The results show that the model pre-trained on our dataset can also perform well on the images captured by different camera sensors by fine-turning. [Best viewed in color.]
}
\label{fig:figure_hdr}
\end{figure*}

\subsection{Generalization to other sensors}
To show the proposed model trained on our dataset can also perform well on other devices/sensors. We choose HDR+ dataset \cite{hasinoff2016burst} as an extra resource. HDR+ is a burst photography dataset for high dynamic range (HDR) and low-light imaging on mobile cameras. The dataset is collected by Android phones along with Android’s Camera2 API. The frame rate in each burst is 15-30 frames per second. Different from our dataset, the aim of HDR+ is for the HDR algorithm originally, the misalignment in a burst is much lower than our dataset. Moreover, the blur is mainly caused by object motion instead of camera motion.

We choose the curated subset which contains 153 bursts. The first five frames of each burst are used to generate blurred images and the middle ones are picked as the sharp ground truths. For the bursts which are less than five frames, we simply discard them. The model we use is our best version in Table \ref{table:ablation_study}, which has pre-trained on our dataset. We set the learning rate at 1e-4 and fine-tune about 90K iterations. The results are shown in Figure \ref{fig:figure_hdr}, and the significant deblurring effect we can observe. With fine-tuning, the model pre-trained on our dataset can also adapt to other sensors.

\section{Discussion and Future Work}
In this work, we emphasize the opportunity for better blurry image synthesis and restoration from the RAW domain. Inspired by Nah et al. \cite{nah2017deep} and Su et al. \cite{su2017deep}, we collect the Deblur-RAW, the first RAW image dataset for deblurring. However, due to the limitation of the current hardware, we can only collect RAW videos with 30 fps. Averaging on low fps videos is prone to generate blurry images with aliasing artifacts, especially for fast-moving objects in the scenes. Prior RGB-based datasets try to address the issue by capturing videos with pretty high fps \cite{nah2017deep} or generating synthetic frames between adjacent frames to increase fps \cite{su2017deep, nah2019ntire}. Although there may be sophisticated cameras with higher fps RAW, to the best of our ability, Canon EOS 6D, the enhanced consumer-level camera, is what we can obtain for RAW video recording. Besides, as the RAW images are stored with the Bayer pattern, prior sRGB frame interpolation methods cannot be directly applied. To the best of our knowledge, there is still no existing method for RAW video frame interpolation. To avoid the aliasing artifacts, we have carefully removed the cases containing fast-moving objects.

Although there are limitations in the Deblur-RAW dataset, it is still the first and promising raw image deblurring dataset. It enables an end-to-end model learning on informative raw sensor data. By a series of experiments, we demonstrate that the image deblurring task can benefit from processing on the RAW domain directly. In our future works, we will address the issue by creating a frame interpolation method for RAW videos that enables us to synthesize high fps RAW videos and alleviate the aliasing issue.

\section{Conclusion}
In this work, we leverage informative RAW images to address the image deblurring problem. As the lack of dataset, previous methods focus on low-bit sRGB image deblurring and lose rich and valuable details which can be gained from RAW images. Therefore, we create Deblur-RAW, the first RAW image deblurring dataset. By the collected dataset, we can directly learn an end-to-end deblurring model from informative raw sensor data. In addition, we propose an innovative network architecture which is more suitable for RAW image deblurring. Through our various and extensive experiments, we demonstrate that the rich and valuable information kept in RAW images benefits the deblurring task. By the proposed dataset and network architecture, we can restore more structure and textural details from RAW images. In conclusion, directly deblurring on raw sensor data differentiates our method from the prior state-of-the-arts and makes us achieve better performance, both quantitatively and qualitatively.

\section*{Acknowledgment}
This work was supported in part by the Ministry of Science and Technology, Taiwan, under Grant MOST 109-2634-F-002-032 and Qualcomm Technologies, Inc. We benefit from NVIDIA DGX-1 AI Supercomputer and are grateful to the National Center for High-performance Computing.

\ifCLASSOPTIONcaptionsoff
  \newpage
\fi



\bibliographystyle{IEEEtran}
\bibliography{IEEEabrv, egbib.bib}
\end{document}